\documentclass[]{aa}

\pdfoutput=1
\usepackage{graphicx}
\usepackage[varg]{txfonts} 
\usepackage{natbib}
\usepackage{subcaption}
\usepackage[breaklinks=true, colorlinks=true, linkcolor=blue, citecolor=blue, urlcolor=blue]{hyperref}
\usepackage{dcolumn}

\begin{document}

\title{Nucleosynthesis in AGB stars traced by oxygen isotopic ratios}
\subtitle{I. Determining the stellar initial mass by means of the $^{17}$O/$^{18}$O ratio}

\author{R.~De~Nutte\inst{\ref{inst1}}
	\and L.~Decin\inst{\ref{inst1},\ref{inst2}}
	\and H.~Olofsson\inst{\ref{inst3}}
	\and R.~Lombaert\inst{\ref{inst1},\ref{inst3}}
	\and A.~de~Koter\inst{\ref{inst1},\ref{inst2}}
	\and A.~Karakas\inst{\ref{inst7},\ref{inst8}}
	\and S.~Milam\inst{\ref{inst4}}
	\and S.~Ramstedt\inst{\ref{inst5}}
	\and R.~J.~Stancliffe\inst{\ref{inst6}}
	\and W.~Homan\inst{\ref{inst1}}
	\and M.~Van~de~Sande\inst{\ref{inst1}}
	}

\institute{Institute of Astronomy, KU Leuven, Celestijnenlaan 200D B2401, 3001 Leuven, Belgium \\
\email{rutger.denutte@ster.kuleuven.be}\label{inst1}
\and 
Astronomical Institute Anton Pannekoek, University of Amsterdam, P.O.~Box 94249, 1090 GE Amsterdam, The Netherlands\label{inst2}
\and
Onsala Space Observatory, Dept. of Earth and Space Sciences, Chalmers University of Technology, 439 92 Onsala, Sweden\label{inst3}
\and
Research School of Astronomy and Astrophysics, Australian National University, Canberra, ACT 2611, Australia
\label{inst7}
\and
Monash Centre for Astrophysics, School of Physics \& Astronomy, Monash University, VIC 3800, Australia
\label{inst8}
\and
NASA Goddard Space Flight Center, Astrochemistry Laboratory, Code 691.0, Greenbelt, MD 20771, USA\label{inst4}
\and
Department of Physics and Astronomy, Uppsala University, Box 516, 751 20 Uppsala, Sweden\label{inst5}
\and
Argelander-Institut f\"{u}r Astronomie, University of Bonn,
Auf dem H\"{u}gel 71, D-53121 Bonn, Germany\label{inst6}
}

\date{Received day month 2016 /
		Accepted day month 2016}

\abstract {}
{The aim of this paper is to investigate the $^{17}$O/$^{18}$O ratio for a sample of AGB stars, containing M-, S- and C-type stars. 
These ratios are evaluated in relation to fundamental stellar evolution parameters: the stellar initial mass and pulsation period.}
{Circumstellar $^{13}$C$^{16}$O, $^{12}$C$^{17}$O and  $^{12}$C$^{18}$O line observations were obtained for a sample of nine stars with various single-dish long-wavelength facilities. 
Line intensity ratios are shown to relate directly to the surface $^{17}$O/$^{18}$O abundance ratio.}
{Stellar evolution models predict the $^{17}$O/$^{18}$O ratio to be a sensitive function of initial mass and to remain constant throughout the entire TP-AGB phase for stars initially less massive than 5\,$M_{\odot}$.
This makes the measured ratio a probe of the initial stellar mass.}
{Observed $^{17}$O/$^{18}$O ratios are found to be well in the range predicted by stellar evolution models that do not consider convective overshooting. From this, accurate initial mass estimates are calculated for seven sources. For the remaining two sources two mass solutions result, though with a larger probability that the low-mass solution is the correct one. Finally, hints at a possible separation between M/S- and C-type stars when comparing the $^{17}$O/$^{18}$O ratio to the stellar pulsation period are presented.}

\keywords{stars: AGB and post-AGB -- stars: evolution -- stars: fundamental parameters -- circumstellar matter} 

\maketitle

\section{Introduction}

Stars with initial mass between 0.8\,$M_{\odot}$ and 8\,$M_{\odot}$ contribute profoundly to the enrichment of the interstellar medium (ISM).
During their passage through the asymptotic giant branch (AGB) these stars lose material through a dense wind driven by radiation pressure on dust grains that form above the photosphere \citep{habing2004agb}.
In order to test and constrain models of AGB evolution it is crucial that the initial mass of these stars can be established.  
However, this has proven to be one of the most elusive problems in stellar astrophysics.  
A promising solution is to use isotopic ratios of elements affected by thermo-nuclear processing, in particular the ratio between $^{17}$O and $^{18}$O.
The chemical composition of these species as they are injected in the ISM is determined by the nucleosynthesis in the stellar center and the subsequent convective-envelope mixing or dredge-ups (DUs). 
Models predict that the first DU changes the surface composition between the main sequence and the tip of the first red giant branch, when convection reaches down into the regions of the star where hydrogen burning has depleted $^{18}$O and enriched $^{17}$O.
As a result the $^{16}$O/$^{17}$O fraction reduces while the $^{16}$O/$^{18}$O fraction marginally increases \citep{boothroyd1994oxygen}. 
Following the end of core helium burning, a second DU event is expected to occur in stars more massive than 4--5\,$M_{\odot}$. 
The effect of this second event is that the surface abundance of $^{16}$O and $^{18}$O only slightly decreases, while $^{17}$O significantly increases.
The effect on the oxygen ratios is the same as for the first DU.
After entering onto the AGB, stars undergo a succession of third DU events, resulting from thermal pulses.
During this phase the surface oxygen isotopic ratios are expected to change only for stars above about 4--5\,M$_{\odot}$ for metallicities close to solar, as a result of hot bottom burning (HBB).
During HBB, the convective envelope penetrates into the H-shell and nuclear reactions at the bottom of the envelope cause the $^{18}$O surface abundance to drop strongly, without affecting the $^{17}$O abundance \citep{lattanzio1997nucleo}.
The impact of these processes on the oxygen isotopic abundances depends on the initial mass of the star, hence it may be used as a probe of this important stellar property. 

\begin{table*}[!ht]
\caption{The sample of AGB stars included in this study, along with relevant stellar and circumstellar parameters. Literature sources and uncertainties of these parameters are discussed in Sect.~\ref{sec:sample}.}
\label{table:sample}
\centering
\begin{tabular}{l l r r r r r r}
\hline\hline 
\noalign{\smallskip}
Source & IRAS & \multicolumn{1}{c}{$D$} & \multicolumn{1}{c}{$P$} & \multicolumn{1}{c}{$L_{\star}$} & \multicolumn{1}{c}{$\dot{M}$} & \multicolumn{1}{c}{$\varv_{\rm{LSR}}$} & \multicolumn{1}{c}{$\varv_\infty$} \\
 & & \multicolumn{1}{c}{[pc]} & \multicolumn{1}{c}{[days]} & \multicolumn{1}{c}{[L$_{\sun}$]} &  \multicolumn{1}{c}{[M$_{\sun}$\,yr$^{-1}$]} &  \multicolumn{1}{c}{[km\,s$^{-1}$]} &  \multicolumn{1}{c}{[km\,s$^{-1}$]} \\
\hline
\noalign{\smallskip}
\multicolumn{4}{l}{\textit{M-stars}} \\
GX Mon & 06500+0829 & 550 & 527 & 8200 & 1.2\,$\times$\,10$^{-5}$ & -9.6 & 19.3 \\
WX Psc & 01037+1219 & 700 & 660 & 10300 & 4.0\,$\times$\,10$^{-5}$ & 9.3 & 18.6 \\
\hline
\noalign{\smallskip}
\multicolumn{4}{l}{\textit{S-stars}} \\
W Aql & 19126-0708 & 300 & 490 & 7600 & 2.7\,$\times$\,10$^{-6}$ & -24.3 & 16.4 \\
$\chi$ Cyg & 19486+3247 & 181 & 407 & 6500 & 6.0\,$\times$\,10$^{-7}$ & 9.0 & 8.3 \\
\hline
\noalign{\smallskip}
\multicolumn{4}{l}{\textit{C-stars}} \\
CW Leo & 09452+1330 & 120 & 630 & 9800 & 1.5\,$\times$\,10$^{-5}$ & -25.9 & 14.2 \\
LL Peg & 23166+1655 & 1300 & 696 & 10900 & 2.5\,$\times$\,10$^{-5}$ & -30.6 & 12.6 \\
LP And & 23320+4316 & 630 & 614 & 9600 & 7.0\,$\times$\,10$^{-6}$ & -16.9 & 13.6 \\
RW LMi & 10131+3049 & 440 & 640 & 10000 & 6.0\,$\times$\,10$^{-6}$ & -2.1 & 16.5 \\
V384 Per & 03229+4721 & 600 & 535 & 8300 & 3.0\,$\times$\,10$^{-6}$ & -16.4 & 14.2 \\
\hline
\end{tabular} 
\end{table*}

The earliest attempts at constraining oxygen isotopic ratios in AGB stars used near-IR observations of the stellar atmospheres. 
In this way \citet{harris1984rgs} and \citet{harris1985ba, harris1985sms, harris1987carbon, harris1988kgiant} determined $^{16}$O/$^{17}$O and $^{16}$O/$^{18}$O ratios for a sample of red giant stars, including barium stars and K giants, five MS- and three S-type stars, and 26 carbon-rich stars. No M-type sources were included in the sample.
For the C-type stars, excluding the J-type stars\footnote{There has been recent evidence that J-type stars might not even be TP-AGB stars at all \citep{sengupta2013jtype}.}, oxygen ratio values of 550 $\leq$ $^{16}$O/$^{17}$O $\leq$ 4100, 700 $\leq$ $^{16}$O/$^{18}$O $\leq$ 2400 and 0.69 $\leq$ $^{17}$O/$^{18}$O $\leq$ 1.56 were found, and 625 $\leq$ $^{16}$O/$^{17}$O $\leq$ 3000, 875 $\leq$ $^{16}$O/$^{18}$O $\leq$ 4700 and 0.78 $\leq$ $^{17}$O/$^{18}$O $\leq$ 2.05 for the MS- and S-type sources.
Note that the method employed in these studies can not be applied to high mass-loss-rate sources, as the high densities in the stellar wind obstruct the view of the atmosphere in the infrared.

The best effort made so far in determining accurate oxygen isotopic ratios from millimeter-wavelength CO observations in the extended circumstellar envelope was done by \citet{kahane1992isotopic} for five carbon-rich envelopes. Isotopic ratios ranging from 242--840 and 317--1260 were found for the $^{16}$O/$^{17}$O and $^{16}$O/$^{18}$O, leading to $^{17}$O/$^{18}$O values between 1.12 and 1.66.
More recent oxygen isotope studies focusing on the circumstellar environment (CSE) were performed by \citet{decin2010iktau} and \citet{khouri2014wind} for the M-type stars IK Tau and W Hya. 
In these works H$_{2}$O isotopologues were used to constrain the isotopic ratios yielding $^{16}$O/$^{17}$O=600$\pm$150, $^{16}$O/$^{18}$O=200$\pm$50 and $^{16}$O/$^{17}$O=$1250_{-450}^{+750}$, $^{16}$O/$^{18}$O=$190_{-90}^{+210}$ respectively. 
The rather low value for the $^{16}$O/$^{18}$O ratio compared to the $^{16}$O/$^{17}$O value leads a $^{17}$O/$^{18}$O ratio of $0.15$ for W Hya, lower than the solar value of $0.1895$ \citep{asplund2009sun}. 
Uncertainties associated with these observational values are large, and do not allow one to draw any definitive conclusions in regard to stellar evolution.
Finally, \citet{justtanont2015ohir} found a high lower limit ($\ge$\,10) to the $^{17}$O/$^{18}$O line intensity ratio in their sample of nine extreme OH/IR stars, consistent with these stars having undergone HBB.

This paper presents the first results in deriving oxygen ratios from circumstellar millimeter-wavelength CO isotopologue observations for a sample of stars covering the three chemical types. 
These ratios are confronted to three independent sets of evolutionary predictions with the aim to constrain their initial masses.
The sample and data reduction process are presented in Sect.~\ref{sec:observations}. 
Section~\ref{sec:results} describes the derivation of the $^{17}$O/$^{18}$O ratios from the line intensities.
These results are linked to stellar evolution models in Sect.~\ref{sec:discussion}.

\section{Observations}
\label{sec:observations}

\subsection{Sample selection}
\label{sec:sample}

The sample for which the $^{17}$O/$^{18}$O ratio was obtained from circumstellar $^{12}$C$^{17}$O and $^{12}$C$^{18}$O J=2-1 and J=1-0 observations comprises nine Mira-type and semi-regular (SRa) AGB sources.  
Miras are relatively well understood in terms of pulsational behavior. 
They are the M-type stars \object{GX\,Mon} and \object{WX\,Psc}; the S-type stars \object{W\,Aql} and \object{$\chi$\,Cyg}, and five C-type stars: \object{LL\,Peg}, \object{CW\,Leo}, \object{LP\, And}, \object{RW\,LMi}, and \object{V384\,Per}. 
The sample thus represents all main chemical types, including M-type stars that so far have not been analyzed with millimeter-wavelength observations in terms of the $^{17}$O/$^{18}$O ratio.  
The stars are relatively nearby sources (within about 1\,kpc) as the inherently weak isotopologue lines of $^{12}$C$^{17}$O and $^{12}$C$^{18}$O are still detectable.  
This sample was analyzed in detail by e.g. \citet{ramstedt2014co}, who derived mass-loss rates and $^{12}$C/$^{13}$C from CO measurements using detailed radiative-transfer models. 

Table~\ref{table:sample} lists all sources, together with distances and relevant stellar and circumstellar properties.
Distances were derived from Hipparcos parallax measurements \citep{vanleeuwen2007hipparcos} but only if the relative error is less than 50\%.
When no accurate Hipparcos data were found, distances were estimated from the period-luminosity ($P$--$L$) relation presented in \citet{groenewegen1996revisedpl}.
The thusly derived luminosities are also listed in Table~\ref{table:sample}.
Periods were taken from the General Catalogue of Variable Stars \citep{samus2009gcvs}.
A conservative error estimate of 10 days was assumed for all stars.
The mass-loss rates $\dot{M}$ were taken from \citet{ramstedt2014co}. 
The Local Standard of Rest velocity $\varv_{\rm{LSR}}$ and the wind terminal velocity $\varv_\infty$ have been calculated from fitting a shell profile to the $^{13}$C$^{16}$O lines, being much brighter than the $^{12}$C$^{17}$O and $^{12}$C$^{18}$O lines. 
The fitted function is 
\begin{equation} \label{eq:shell}
f(\nu)=\frac{A}{\delta\nu}\,\frac{1+4H\left[(\nu-\nu_0)/\delta\nu\right]^2}{1+H/3},
\end{equation} 
where $A$ is the area under the profile, $\delta\nu$ is the full width at zero level, and $\nu_0$ is the central frequency \citep{CLASSsite}.
The Horn/Center parameter $H$ dictates the shape of the function, as
\begin{equation} \label{eq:shellH}
\frac{f(\delta\nu/2)}{f(0)}=1+H.
\end{equation} 

\subsection{Millimeter-wavelength observations and data reduction}

The data presented in this study have been obtained with three different telescopes. The Institut de Radio Astronomie Millimétrique (IRAM) 30m telescope \citep{baars1987IRAM} at Pico Veleta, Spain (Program ID 042-12, 164-12) using the EMIR heterodyne receiver in dual band observation mode in the E0(90 GHz)/E2(230 GHz) configuration; the Atacama Pathfinder EXperiment (APEX) 12m telescope \citep{gusten2006APEX} on the Chajnantor Plateau, Chile (Programme ID 090.D-0290, 091.D-0813, 094.D-0851A), in the SHeFI 230 GHz band, and the Caltech Submillimeter Observatory (CSO) at Mauna Kea, Hawaii, using the 230 GHz receiver.
The sources were observed using position switching or wobbler switching mode to attain flat baselines.
The pointing of the telescope was checked repeatedly throughout the observations using strong CO and continuum sources.
The reduction and analysis of all data were performed using the GILDAS CLASS software package \citep{CLASSsite}.
After removing faulty scans and spikes, a first-order polynomial baseline was subtracted from each scan. 
The individual baseline-subtracted scans obtained for a given source were then averaged using an inverse quadratic system temperature weighting, with weighting factor
\begin{equation} 
	\label{eq:weights}
	w_{i}\,=\,\frac{\Delta t\Delta \nu}{T_{\rm{sys}}^{2}},
\end{equation} 
where $\Delta t$ and $\Delta \nu$ are the integration time and frequency resolution respectively. 
Finally the data were rebinned to obtain a suitable signal-to-noise ratio (SNR), generally SNR $\approx$ 3--5 for a velocity resolution of 2 km/s (the typical line width being around 20--40\,km/s).
A conversion from observed source antenna temperature $T_{A}^{\star}$ (corrected for atmospheric and radiative loss, and rearward scattering and spillover) to main-beam-temperature scale $T_{\rm{mb}}$ through 
\begin{equation} \label{eq:Tmb}
T_{\rm{mb}}\,=\,\frac{T_{A}^{\star}}{\eta_{\rm{mb}}}
\end{equation} 
was performed, where $\eta_{\rm{mb}}$ is the telescope main-beam efficiency. 
The main-beam efficiency for a given wavelength can be interpolated from values measured in \citet{gusten2006APEX} for APEX, \citet{magnum1993CSO} for CSO, and \citet{baars1987IRAM} and \citet{kramer2013improvement} for IRAM. 
The telescope beam sizes $\theta_{\rm{mb}}$ are wavelength-dependent and can be calculated through $\theta_{\rm{mb}}\,=\,k\lambda/D$ [rad] with $k$ ($1\leqslant k\lesssim 1.4$) a telescope-specific factor and $D$ the telescope main dish size. 

Table \ref{table:lines} lists the measured integrated line intensities and maximum main beam brightness temperatures for all detected $^{12}$C$^{17}$O and $^{12}$C$^{18}$O lines, which are also shown in Fig.~\ref{fig:profiles}.

\begin{table}[!ht]
\caption{Observational results for the CO isotopologue line detections. Detected lines are labeled by a three-part code representing the isotopologue (17: $^{12}$C$^{17}$O, 18: $^{12}$C$^{18}$O), telescope used (A: APEX, C: CSO, I: IRAM) and transition (e.g. 21 being the J=2-1 transition).}
\label{table:lines}
\centering
\begin{tabular}{l l c c c}
\hline\hline 
\noalign{\smallskip}
Source & Line & $\nu_{trans}$ & $I_{\rm{mb}}$ & $T_{\rm{mb}}$ \\
 &  & [GHz] & [K\,km\,s$^-1$] & [mK] \\
\hline
\noalign{\smallskip}
\multicolumn{4}{l}{\textit{M-stars}} \\
GX Mon & 17I21 & 224.714 & 0.71 & 31 \\
 & 18I21 & 219.560 & 0.88 & 45 \\
WX Psc & 17A21 & 224.714 & 0.16 & 11 \\
 & 18A21 & 219.560 & 0.60 & 15 \\
\hline
\noalign{\smallskip}
\multicolumn{4}{l}{\textit{S-stars}} \\
W Aql & 17A21 & 224.714 & 0.38 & 19 \\
 & 18A21 & 219.560 & 0.31 & 13 \\
$\chi$ Cyg & 17I10 & 112.360 & 0.12 & 9.5 \\
 & 18I10 & 109.782 & 0.053 & 2.7 \\
 & 17I21 & 224.714 & 0.24 & 19 \\
 & 18I21 & 219.560 & 0.13 & 10 \\
\hline
\noalign{\smallskip}
\multicolumn{4}{l}{\textit{C-stars}} \\
CW Leo & 17C21 & 224.714 & 1.72 & 103 \\
 & 18C21 & 219.560 & 1.38 & 121 \\
LL Peg & 17A21 & 224.714 & 0.39 & 21  \\
 & 17I21 & 224.714 & 0.70 & 30 \\
 & 18A21 & 219.560 & 0.72 & 39 \\
 & 18I21 & 219.560 & 1.11 & 46 \\
LP And & 17I21 & 224.714 & 1.39 & 60 \\
 & 18I21 & 219.560 & 0.98 & 48 \\
RW LMi & 17I21 & 224.714 & 1.39 & 52 \\
 & 18I21 & 219.560 & 0.98 & 38 \\
V384 Per & 17I21 & 224.714 & 0.71 & 29 \\
 & 18I21 & 219.560 & 0.33 & 13 \\
\hline
\end{tabular} 
\vspace{-3mm}

\end{table}

\section{Results}
\label{sec:results}

\subsection{Deriving $^{17}{\rm O}$\,/\,$^{18}{\rm O}$ ratios}

In determining isotopic ratios from observations of extended circumstellar envelopes one must take into account the isotope-selective nature of astrochemical processes, such as photodissociation. 
Because of the lower abundance of the rare CO isotopologues, line self-shielding for $^{13}$C$^{16}$O, $^{12}$C$^{17}$O and $^{12}$C$^{18}$O is much less efficient than for $^{12}$C$^{16}$O.
For $^{13}$C$^{16}$O this effect has been shown to be countered by chemical fractionation \citep{mamon1988photodissociation}.
As the photodissociation rates for $^{12}$C$^{17}$O and $^{12}$C$^{18}$O are expected to have an almost identical radial depth dependence \citep{visser2009photodissociation}, these species may allow for a robust determination of the intrinsic $^{17}$O/$^{18}$O ratio.

Calculating the intrinsic $^{17}$O/$^{18}$O abundance ratios from  $^{12}$C$^{17}$O and $^{12}$C$^{18}$O line intensities can be done in a fairly straightforward way.
Firstly, a correction must be made for the different Einstein coefficients and beam widths which, combined, lead to 
\begin{equation} \label{eq:freqcorr}
^{17}\rm{O}/^{18}\rm{O}\,=\,\frac{I_{\rm{mb}} \Big({}^{12}\rm{C}^{17}\rm{O} \left(J \rightarrow J-1 \right)\Big)}{I_{\rm{mb}} \Big({}^{12}\rm{C}^{18}\rm{O} \left(J \rightarrow J-1 \right)\Big)} \times \left(\frac{\nu_{^{12}\rm{C}^{17}\rm{O} \left(J \rightarrow J-1 \right)}}{\nu_{^{12}\rm{C}^{18}\rm{O} \left(J \rightarrow J-1 \right)}}\right)^{-3},
\end{equation}
with $\nu$ the frequency of the considered transition and $I_{\rm{mb}}$ the measured integrated line intensity.
For all transitions considered, the frequency ratio correction factor in Eq. \ref{eq:freqcorr} is equal to 0.933.
For those sources for which more than one pair of lines were obtained, an average of the line intensity ratios weighted by the uncertainties on the line strength was adopted.

\subsection{Uncertainties on the calculated ratios}

Optical depth effects, if present, will lead to a non-linear relation between line intensity and molecular column density.
In order to determine the extent of these effects, a grid of model circumstellar envelopes was computed for an appropriate range of
$L_{\star}$, $\dot{M}$, C/O and $^{17}$O/$^{18}$O input abundances
using the non-local thermodynamic equilibrium (NLTE) radiative-transfer code GASTRoNOoM \citep{decin2006probing, decin2010circumstellar, lombaert2013h2o}.
Required optical-depth corrections on the observational $^{17}$O/$^{18}$O ratios as determined through Eq. \ref{eq:freqcorr} were then computed from this grid and were found to be on the order of $\sim$5\% at most.
This is indeed in line with these lines being optically thin and thus supports the notion that the observational line intensity ratios can indeed be directly related to the intrinsic $^{17}$O/$^{18}$O ratios.

Table~\ref{table:ratios} lists the observed $^{17}$O/$^{18}$O ratios and their respective uncertainties.
These uncertainties are calculated through standard error propagation on the integrated line strength uncertainties, which are themselves determined by the root mean square fluctuations per spectral channel aggregated over the line width.
The uncertainty on the averaged $^{17}$O/$^{18}$O ratio is lowered for sources for which more than one pair of lines was detected.
As the $^{12}$C$^{17}$O and $^{12}$C$^{18}$O lines were always measured consecutively in the same instrument band, the calibration uncertainty is significantly lowered when calculating the ratio of these line strengths (from typically 20--30\% uncertainty on such weak lines to about a 5\% relative accuracy between the lines).

\begin{table*}[!htb]
\caption{Observed $^{17}$O/$^{18}$O ratios and their uncertainties. The final three columns denote the initial-mass estimates calculated from these ratios through comparison with stellar-evolution models.}
\label{table:ratios}
\setlength{\tabcolsep}{4pt}
\centering
\begin{tabular}{l c c c c}
\hline\hline 
\noalign{\smallskip}
Source & $^{17}$O/$^{18}$O & $M_{\rm{i}}$ \citep{stancliffe2004code} & $M_{\rm{i}}$ \citep{karakas2016yields2} & $M_{\rm{i}}$ \citep{cristallo2011fruity} \\
 & & [$M_{\odot}$] & [$M_{\odot}$] & [$M_{\odot}$] \\
\hline
\noalign{\smallskip}
\multicolumn{4}{l}{\textit{M-stars}} \\
GX Mon & 0.77\,$\pm$\,0.16 & $1.43^{+0.07}_{-0.08}$ & $1.49^{+0.04}_{-0.09}$ & $1.52^{+0.04}_{-0.07}$\\[2pt]
WX Psc & 0.26\,$\pm$\,0.06 & $1.06^{+0.10}_{-0.10}$ & $0.8 \leq M_{\rm{i}} \leq 1.0$ & x \\[2pt]
\hline
\noalign{\smallskip}
\multicolumn{4}{l}{\textit{S-stars}} \\
W Aql & 1.17\,$\pm$\,0.22 & $1.57^{+0.06}_{-0.06}$ & $1.61^{+0.07}_{-0.07}$ & $1.62^{+0.06}_{-0.06}$ \\[2pt]
$\chi$ Cyg & 2.00\,$\pm$\,0.52 & $1.79^{+0.14}_{-0.14}$  / $3.69^{+2.31}_{-0.98}$ & $1.83^{+0.13}_{-0.14}$  / $3.69^{+2.31}_{-0.88}$ & $1.84^{+0.14}_{-0.14}$  / $4.68^{+1.32}_{-1.53}$ \\[2pt]
\hline
\noalign{\smallskip}
\multicolumn{4}{l}{\textit{C-stars}} \\
CW Leo 	& 1.16$\pm$0.06 & $1.57^{+0.02}_{-0.02}$ & $1.60^{+0.02}_{-0.02}$ & $1.62^{+0.02}_{-0.02}$ \\[2pt]
LL Peg 	& 0.56$\pm$0.06 & $1.34^{+0.03}_{-0.03}$ & $1.37^{+0.04}_{-0.04}$ & $1.42^{+0.04}_{-0.04}$ \\[2pt]
LP And 	& 1.26$\pm$0.08 & $1.60^{+0.03}_{-0.03}$ & $1.63^{+0.03}_{-0.03}$ & $1.65^{+0.02}_{-0.02}$ \\[2pt]
RW LMi 	& 1.36$\pm$0.19 & $1.62^{+0.06}_{-0.06}$ & $1.66^{+0.06}_{-0.06}$ & $1.67^{+0.05}_{-0.05}$ \\[2pt]
V384 Per 	& 2.04$\pm$0.32 & $1.80^{+0.09}_{-0.09}$ / $3.59^{+0.71}_{-0.71}$ & $1.83^{+0.08}_{-0.08}$ / $3.64^{+0.89}_{-0.74}$ & $1.85^{+0.09}_{-0.09}$ / $4.44^{+1.05}_{-0.96}$ \\[2pt]
\hline
\end{tabular} 
\end{table*}

\section{Discussion}
\label{sec:discussion}

\subsection{Initial stellar mass determination through the $^{17}{\rm O}$\,/\,$^{18}{\rm O}$ ratio}
\label{subsec:mass}

\begin{figure}[!t]
\resizebox{\hsize}{!}{\includegraphics{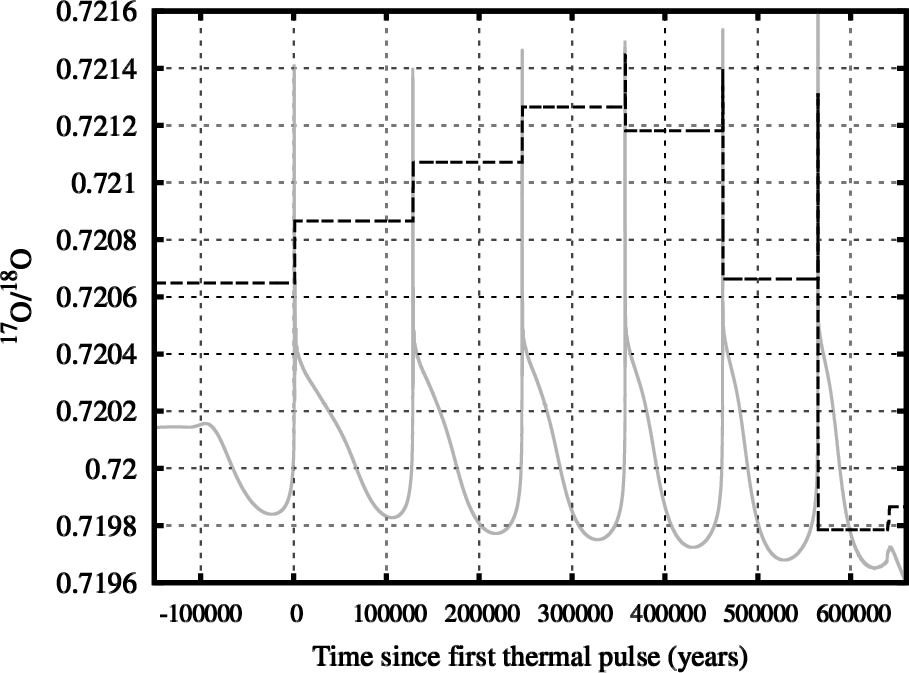}}
\caption{The $^{17}$O/$^{18}$O ratio (dashed black line) for a model star of 1.5\,M$_{\odot}$ and metallicity $Z=0.02$ as a function of time since the first thermal pulse in the TP-AGB phase as obtained with the STARS stellar evolution code presented in \citet{eggleton1971code,stancliffe2004code,stancliffe2009code2}. The helium burning luminosity is overplotted (solid grey line) to mark the various thermal pulses. This model does not include any extra mixing or convective overshooting. Note that the $^{17}$O/$^{18}$O ratio varies insignificantly over the entire TP-AGB phase.}
\label{fig:modelRichard}
\end{figure}

\begin{figure}[!t]
\resizebox{\hsize}{!}{\includegraphics{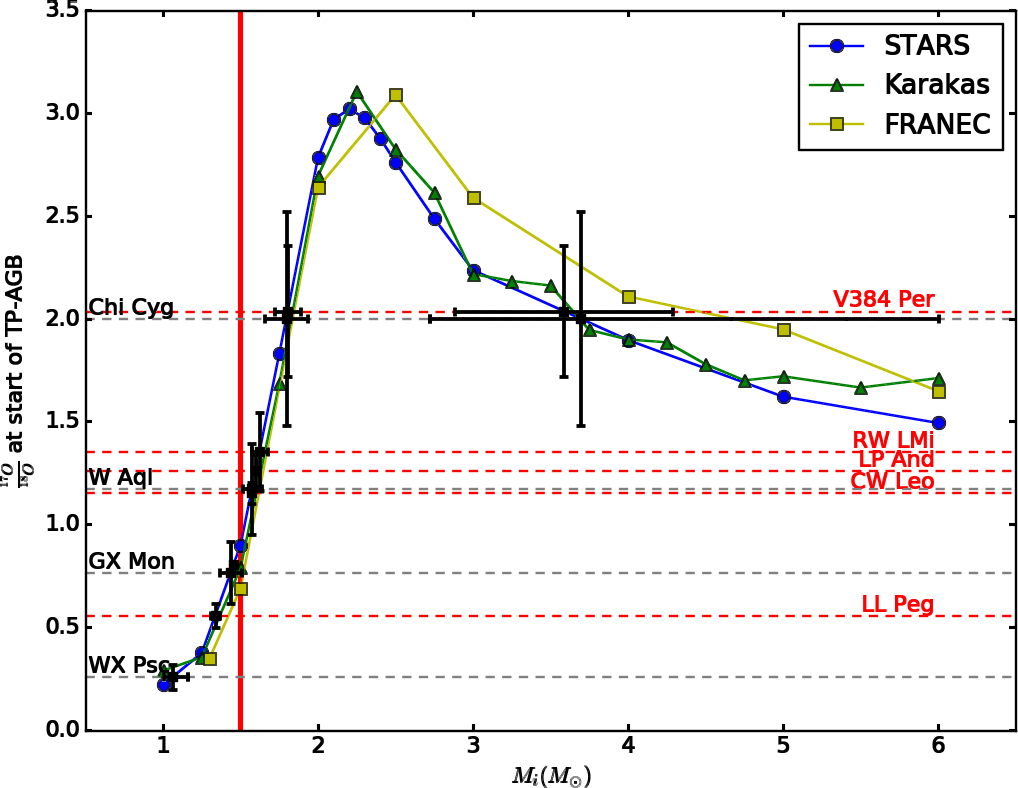}}
\caption{Observational $^{17}$O/$^{18}$O ratios (horizontal lines) compared to values obtained from various stellar evolution models. Black and red horizontal lines represent M/S- and C-type stars respectively. The stellar evolution models are computed with the STARS code \citep{eggleton1971code,stancliffe2004code,stancliffe2009code2}, the FRANEC code \citep{cristallo2011fruity}, and the Stromlo/Monash code \citep{lattanzio1986code,karakas2010yields,karakas2014code}. The red vertical line marks the soft limit for the formation of a carbon-rich star at 1.5\,$M_{\sun}$. The black dots show the projections of the observed $^{17}$O/$^{18}$O ratios (and their uncertainties) onto the interpolated predictions of the STARS evolution code, used to derive the initial masses of the targets.}
\label{fig:compare}
\end{figure}

For given initial mass $M_{\rm i}$ and metallicity $Z$, the surface abundances of isotopes may be anticipated to depend on the number of dredge-ups the star has experienced, on the abundance of the primary element in the inter-shell zone \citep{karakas2010extra} and the stellar core mass \citep{kahane2000improved}.  
However, stellar evolution models (see Fig.~\ref{fig:modelRichard}) show that once the star enters the TP-AGB phase, the $^{17}$O/$^{18}$O ratio remains essentially constant regardless of the number of TPs, barring the more massive stars which undergo HBB.  
So, if the metallicity is known, one may constrain the initial stellar mass directly from the $^{17}$O/$^{18}$O ratio. 
This is done here assuming a solar metallicity for the entire sample. 

To illustrate the principle, Fig.~\ref{fig:compare} shows the methodology using models computed with various stellar evolution codes for a solar metallicity $Z=0.014$ and a negligible convective core overshooting. 
These codes are the STARS code \citep{eggleton1971code,stancliffe2004code,stancliffe2009code2}, the FRANEC code \citep{cristallo2011fruity} and the models by \citet{karakas2016yields2}. 
Two key remarks can be made regarding this comparison between observed and predicted ratios. 
The observed $^{17}$O/$^{18}$O ratios are in the range as predicted by the stellar evolution models.  
This is in line with a similar finding by \citet{lebzelter2015giants} for red giants native to open clusters with known turn-off mass. 
For three of the four stars in their sample the empirical $^{17}$O/$^{18}$O was, within uncertainties, in agreement with values predicted with the stellar evolution models; for one source the empirical value was lower than the theoretical one.
The three independently developed stellar evolution models compared here agree in the $^{17}$O/$^{18}$O ratio prediction. 
This suggests that the initial mass of AGB stars may be constrained with confidence using this method (but see below for a stipulation).

The predictions show a maximum of $^{17}$O/$^{18}$O at an initial mass around 2.5\,$M_{\odot}$. 
The cause hereof is that the more massive the star, the deeper its first dredge-up will reach.  
This will increase the $^{17}$O/$^{18}$O ratio, explaining the initial rise of the curve. 
More massive stars undergo non-degenerate helium ignition, which causes the envelope to retreat and the star to shrink.  
This lessens the dredge-up effect and in turn lowers the isotopic ratio, explaining why the predicted $^{17}$O/$^{18}$O curves in Fig.~\ref{fig:compare} feature a maximum.
This local maximum causes an ambiguity in using this graph to estimate the initial mass from the observed $^{17}$O/$^{18}$O ratio, when two intercepts between stellar evolution predictions and observational ratios are found. 
The constraints on the initial masses for the sample of stars are given in Table~\ref{table:ratios}.

Marked in Fig.~\ref{fig:compare} is the lowest initial mass (1.5\,$M_{\sun}$) for which models can evolve to a carbon star.
This lower limit on the initial mass was also derived observationally by \citet{groenewegen1995carbon} for galactic C-type stars in binary systems and open clusters.
The initial mass found for the C-type star LL Peg is however found to be slightly lower than this value.
As discussed in \citet{stancliffe2005lmc} it seems apparent from observations of carbon star luminosities in the Magellanic clouds that efficient dredge-up should occur at lower core masses than is predicted by current models.
If this finding can be extrapolated to a galactic environment, it would imply a shift of the red line in Fig.~\ref{fig:compare} to lower mass values, likely resolving this inconsistency.

\begin{figure}[!b]
\resizebox{\hsize}{!}{\includegraphics{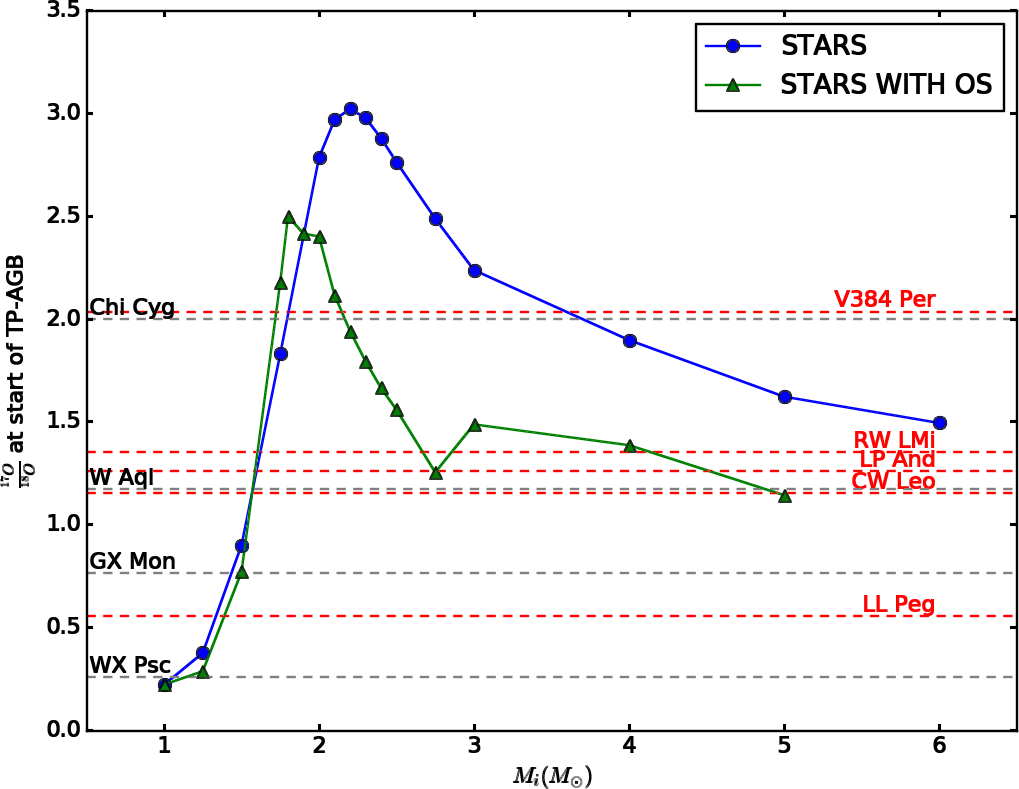}}
\caption{Observational $^{17}$O/$^{18}$O ratios (horizontal lines) compared to values obtained from the STARS stellar evolution code for both the case that  convective core overshooting is neglected (as in Fig.~\ref{fig:compare}) and for an overshooting parameter $\delta_{\rm ov} = 0.156$, following \citep{stancliffe2015os}.}
\label{fig:OScompare}
\end{figure}

Finally, it should be noted that no convective overshooting was considered in the stellar models. 
The extent of core overshooting in main sequence  stars is still uncertain \citep[see e.g.][]{stancliffe2015os,moravveji2015os,claret2016os}.
Because core overshooting increases the mass of the core, the effect of its inclusion is to make a given star behave as though it had a higher initial mass. 
Fig.~\ref{fig:OScompare} shows the results of an additional set of models calculated with the STARS code, which includes convective overshooting using the calibration provided by \citet{stancliffe2015os}, i.e., an overshooting parameter $\delta_{\rm ov} = 0.156$. 
The effect of the inclusion of overshooting is to reduce the $^{17}$O/$^{18}$O ratio in stars of above around 2\,$M_{\odot}$, producing a much narrower peak in the $^{17}$O/$^{18}$O ratio as a function of mass (see Fig.~\ref{fig:OScompare}). 
The maximum $^{17}$O/$^{18}$O ratio is reached at 1.8\,$M_{\odot}$ and the rise to this maximum is only slightly steeper than in the standard case. 
Hence, one may conclude that mass estimates up to 1.8\,$M_{\odot}$ are not significantly affected by uncertainties in the treatment of convective core overshooting.

\subsection{Discriminating between high and low mass estimates}
\label{sec:low_high_mass}

\begin{figure*}[!ht]
\resizebox{\hsize}{!}{\includegraphics{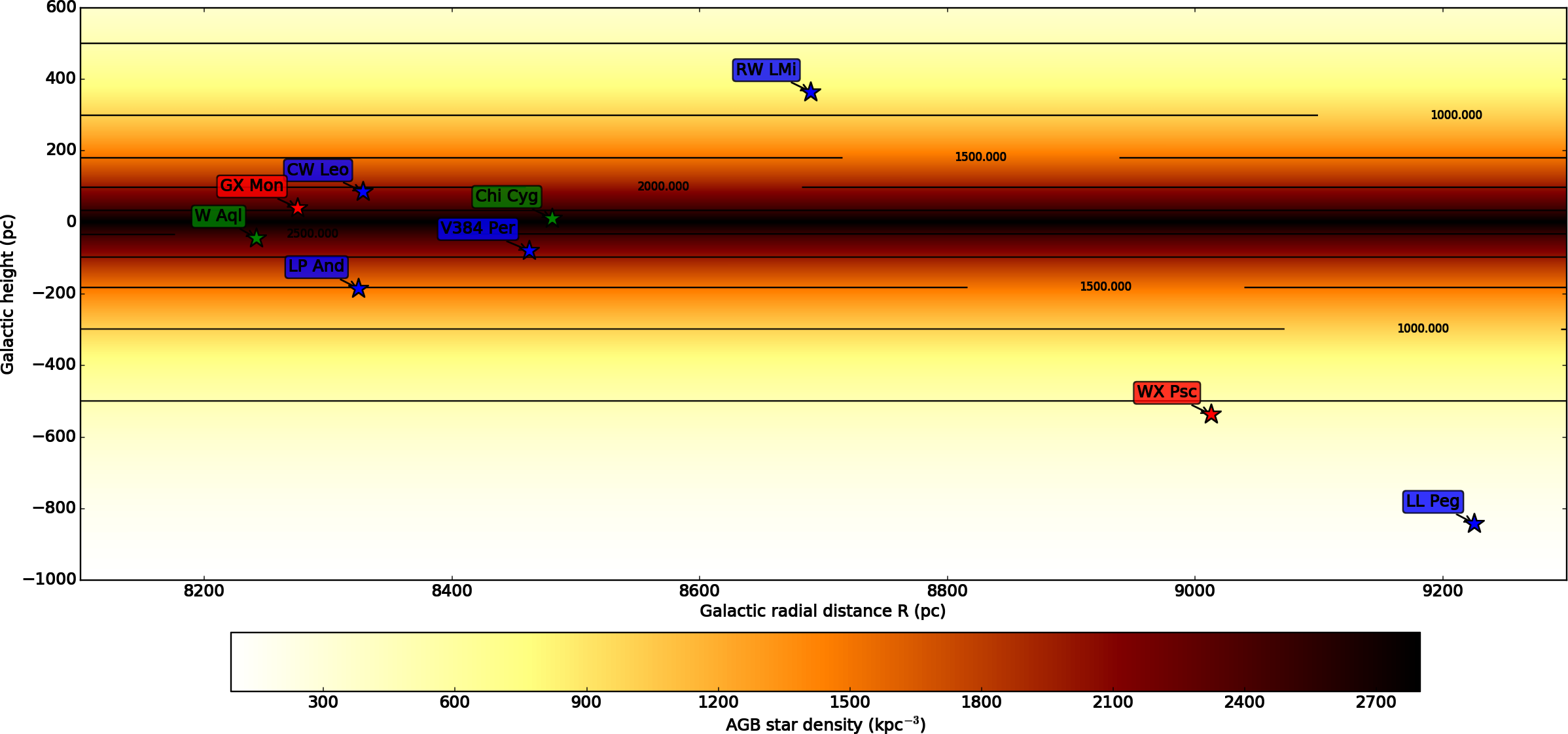}}
\caption{The galactic height and radial distance to the galactic center for the sample of AGB stars. Red, green and blue data points represent M-, S- and C-type stars respectively. Overplotted is the galactic star density distribution of AGB stars as determined in \citet{jackson2002distribution}.}
\label{fig:galdist}
\end{figure*}

\begin{figure}[!hb]
\resizebox{\hsize}{!}{\includegraphics{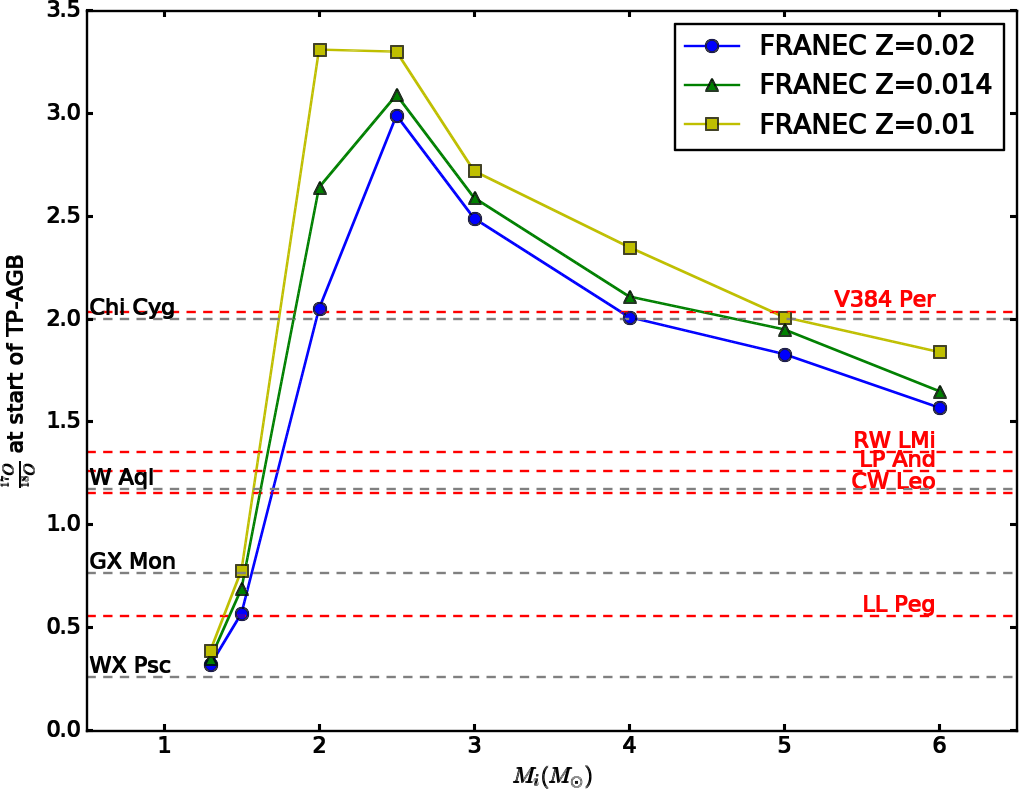}}
\caption{$^{17}$O/$^{18}$O ratios obtained from the FRANEC code \citep{cristallo2011fruity} for different metallicities: solar (${Z=0.014}$, as previously assumed for deriving initial masses) and slightly higher ($Z=0.02$) and lower ($Z=0.01$) metallicity. }
\label{fig:metal}
\end{figure}

For V384\,Per and $\chi$\,Cyg the measured $^{17}$O/$^{18}$O ratio is so large that two solutions for the initial mass are possible, a low mass solution and a high mass solution.  
One may use additional constraints to estimate the likelihood of each solution; two of which are briefly discussed here.

The initial mass function (IMF) favors the formation of low mass stars compared to high mass stars.
For a Salpeter IMF, $n(M)\,\sim\,M^{-2.35}$ \citep{salpeter1955imf}, one can readily calculate that solely on the basis of this argument the lower mass estimates are a factor $\sim$\,4--7 more likely than the high mass estimates.  
One should however realize that the duration of the thermally-pulsing AGB phase for lower mass stars is less than for higher mass stars, an effect that may amount to a factor of  two \citep{rosenfield2016evolution}.
The distribution of the sample of stars with respect to the galactic disk can also be used to put constraints on the initial mass.
Fig.~\ref{fig:galdist} shows this distribution for the entire sample and compares this to the galactic AGB density function as determined by \citet{jackson2002distribution}. 
The galactic disk scale height has been shown to vary for stars of different initial masses \citep[Chapter~2]{sparke2007galaxies}.
The distribution of higher mass (younger) stars is found to be more concentrated toward the galactic plane compared to the population of lower mass (older) stars.
This favors a lower mass for V384\,Per, which is located 80\,pc below the galactic plane. Taken all these arguments together, one may conclude that the high-mass solutions presented in Table \ref{table:ratios} are less probable by a factor of a few.

One may also note the relatively large galactic height of the C-star LL Peg in Fig.~\ref{fig:galdist}. 
This may argue against a solar metallicity for this source. 
Stars with a lower metal content are expected to become C-stars at lower initial mass, which would be in line with the estimate of its mass being 1.34--1.42 M$_{\odot}$.

\subsection{Effect of a non-solar metallicity}

As shown in \citet{cheng2012metallicity}, the radial metallicity gradient at low galactic height (below 250 pc), as is the case for most of the stars in the sample (see Fig.~\ref{fig:galdist}), is about -0.066 dex\,kpc$^{-1}$ in [Fe/H].
This amounts to a change in $Z$ of about 15\% at a radial distance of 1 kpc from the Sun.
This effect therefore does not grossly invalidate the assumption of a solar metallicity for the stars in the sample.

Fig.~\ref{fig:metal} shows the impact of a modest change in metallicity on the modeled $^{17}$O/$^{18}$O ratios.
Predictions for a larger range of metallicities may be found in \citet{karakas2016yields2}.
A lower metallicity will lead to larger predicted $^{17}$O/$^{18}$O ratios overall.
However, due to the steep slope at low initial masses, this leads to a rather small difference in the initial-mass estimate.
At the high-mass end of the graph, the modeled change in metallicity may lead to a discrepancy in derived initial mass of about 1 M$_{\odot}$.
The two sources for which also a high mass solution was found, V384 Per and $\chi$ Cyg, have a galactic radial distance very close to that of the Sun.
Therefore, {\em if} they would have a relatively high mass (but see Sect.~\ref{sec:low_high_mass}) then the mass estimates given in Table~\ref{table:ratios} likely do not suffer from sizable systematic uncertainties due to a potential metallicity effect.

\subsection{Period dependency of the $^{17}$O/$^{18}$O ratio}
\label{subsec:period}

\begin{figure}[!tb]
\resizebox{\hsize}{!}{\includegraphics{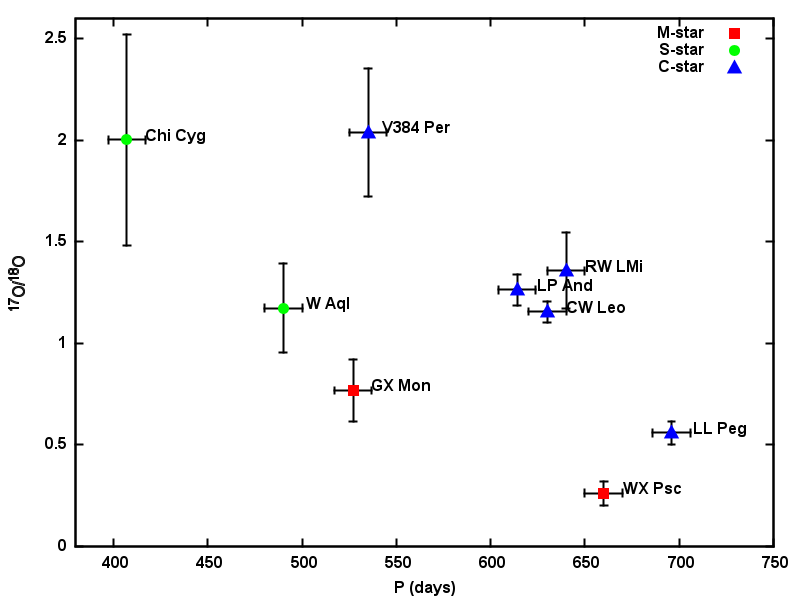}}
\caption{Acquired $^{17}$O/$^{18}$O values plotted versus pulsation period. Data points mark the ratios calculated as described in Sect.~\ref{sec:results} with different markers for C-, M- and S-stars, along with the respective uncertainties.}
\label{fig:17O18O_vs_P}
\end{figure}

\begin{figure}[!tb]
\resizebox{\hsize}{!}{\includegraphics[angle=-90]{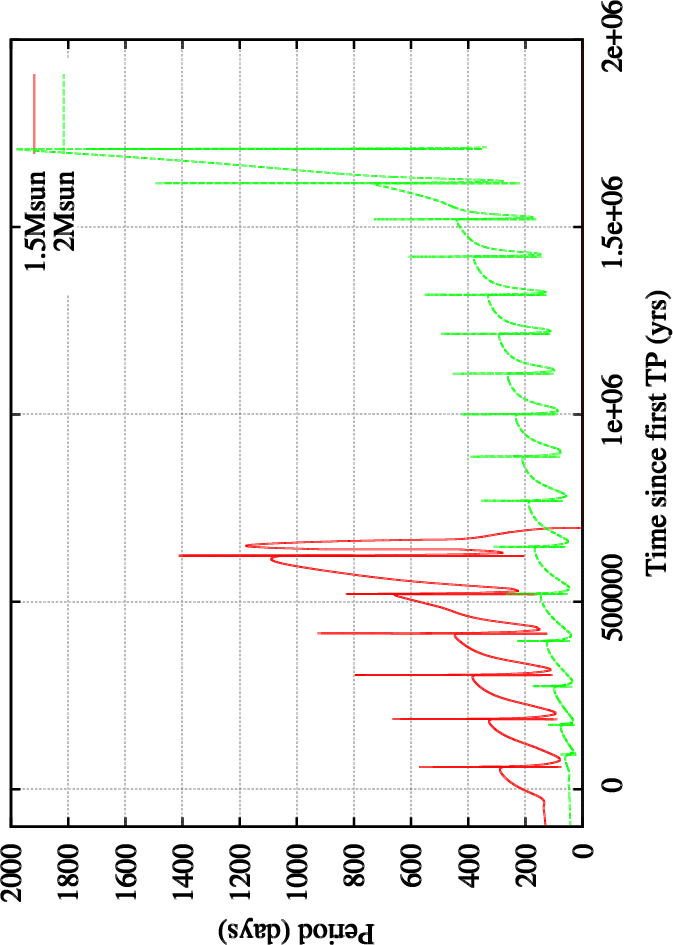}}
\caption{Variation of the stellar pulsation period for a thermally-pulsating star of 1.5\,M$_{\odot}$ and 2\,M$_{\odot}$ during its lifetime, as determined from stellar evolution models.}
\label{fig:periodRichard}
\end{figure}

Fig.~\ref{fig:17O18O_vs_P} shows the $^{17}$O/$^{18}$O ratio as a function of pulsation period. 
As the isotopic ratio is not expected to change during the thermally-pulsing AGB phase, stars are expected to evolve horizontally in this diagram — $^{17}$O/$^{18}$O reflecting initial mass and the stellar period directly indicating stellar lifetime (as shown in Fig.~\ref{fig:periodRichard}).  
\citet{wood1981period} show that the pulsation period $P \sim Q\,R^{\alpha}\,M_{\rm p}^{-\beta}$, where $Q$ is a pulsation constant, $R$ is the stellar radius, and $M_{\rm p}$ is the present mass.   
Values for $\alpha$ range between 1.5$-$2.5 and those for $\beta$ between 0.5$-$1.0, depending on pulsation mode. 
Given this dependence, one may expect that for constant initial mass the evolution is from the left to the right. 
Hence, C-stars may be anticipated to be to the right of the M-type stars, being on average slightly more luminous and somewhat cooler, hence having larger radii. 
This effect is indeed visible in Fig.~\ref{fig:17O18O_vs_P}.

The pulsation constant also modestly depends on molecular opacities \citep{fox1982pulsconst}.  
Potentially, the horizontal separation (at constant isotopic ratio) between M- and S-stars on the one hand and C-stars on the other hand might therefore also (at least in part) reflect a difference in the value of the pulsation constant.  
A significantly larger $Q$ for C-stars is, however, not expected (P. Wood, priv. comm.).

The figure might give the impression of a downward linear trend of the $^{17}$O/$^{18}$O ratio with period, for both the group of M- and S-stars and that of C-stars. 
No such trend is obviously implied by evolutionary predictions, and may be spurious -- reflecting small number statistics or a selection bias.
Employing a statistical analysis on a model distribution of stars, one finds that the probability of four randomly picked stars being on the fit of M- and S-stars is 19.6\% and that of five stars being on the fit of C-stars is 9.3\%.
Up to seven stars would need to be aligned on either of the two fits before one may exclude at three-sigma level the hypothesis that the stars have been randomly picked from the distribution in the $P-^{17}$O/$^{18}$O plane.

The authors aim to further investigate this apparent trend by increasing the sample size and using other molecules tracing the $^{17}$O/$^{18}$O ratio, thus evading the small-number statistics inherent to the current sample size.
In addition, any possible selection bias will be studied through a stellar population synthesis analysis.

\section{Conclusions}

A study of the $^{17}$O/$^{18}$O ratio was performed for a total of nine AGB stars, of which five are carbon stars and four M/S-type stars. 
The intrinsic $^{17}$O/$^{18}$O abundance ratios for these sources were derived from newly obtained circumstellar millimeter-wavelength CO isotopologues observations.
Using stellar evolution models this ratio was shown to be a sensitive tracer of the initial stellar mass.
This allowed for an accurate determination of initial mass estimates for seven of the AGB stars. For two stars, $\chi$\,Cyg and V384\,Per, two solutions for the
mass were obtained with the low-mass solution being most likely.
A separation between carbon stars and M/S-type stars was seen to emerge when investigating the $^{17}$O/$^{18}$O ratio versus the stellar period.
It is clear that a larger sample study is required to attain a statistically relevant conclusion on this matter.

\begin{acknowledgements}
This publication is based on data acquired with the Atacama Pathfinder Experiment (APEX) telescope, the IRAM 30m telescope, and the Caltech Submillimeter Observatory (CSO) telescope.
APEX is a collaboration between the Max-Planck-Institut fur Radioastronomie, the European Southern Observatory (ESO), and the Swedish National Facility for Radio Astronomy, Onsala Space Observatory (OSO). 
IRAM is supported by INSU/CNRS (France), MPG (Germany) and IGN (Spain). 
CSO was operated by the California Institute of Technology under cooperative agreement with the National Science Foundation (AST-0838261). 
This work has benefited from funding from the European Community's Seventh Framework Programme.
LD acknowledges support from the ERC consolidator grant 646758 AEROSOL and the FWO Research Project grant G024112N.
HO acknowledges financial support from the Swedish Research Council.
Finally, the authors would like to thank the anonymous referee for their constructive comments.
\end{acknowledgements}

\bibliographystyle{aa}
\bibliography{refs}

\listofobjects

\Online

\begin{appendix}
\onecolumn

\section{Supporting material}

\vspace{-2mm}

\begin{figure*}[!htb]
\centering
\begin{tabular}{cc}
\includegraphics[width=.41\textwidth]{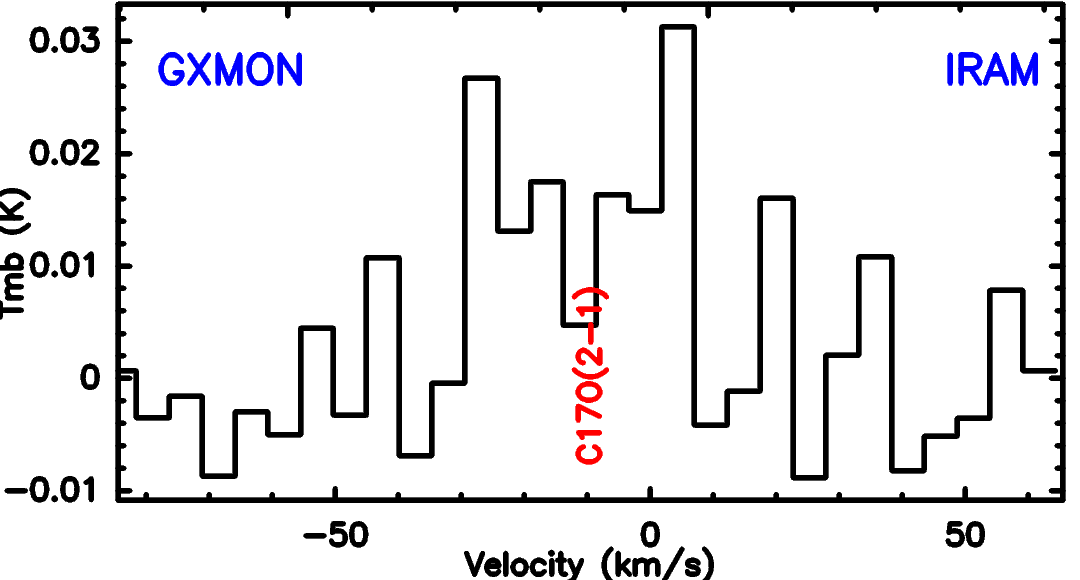} &
\includegraphics[width=.41\textwidth]{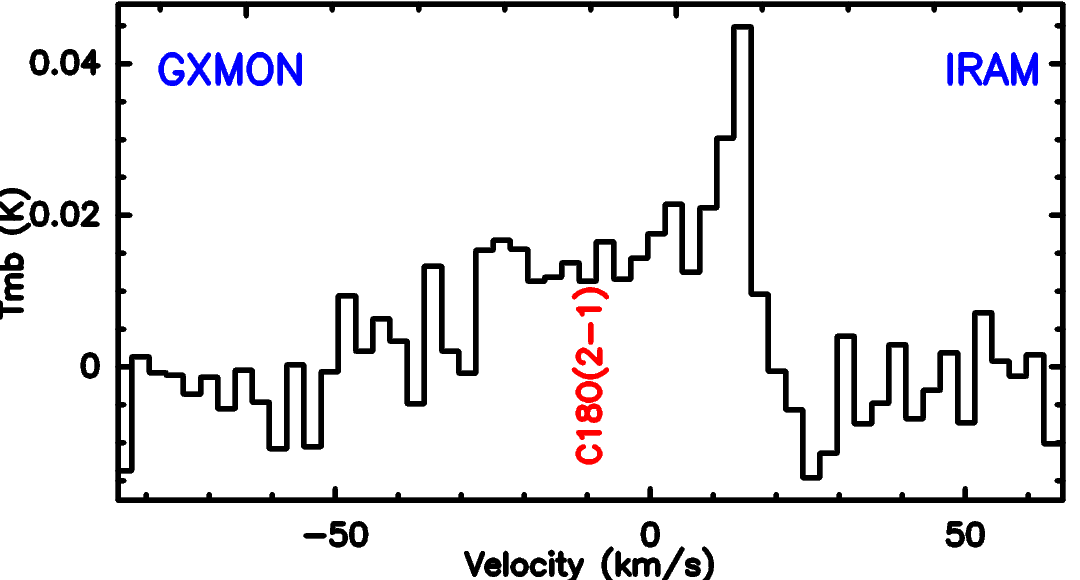} \\
\includegraphics[width=.41\textwidth]{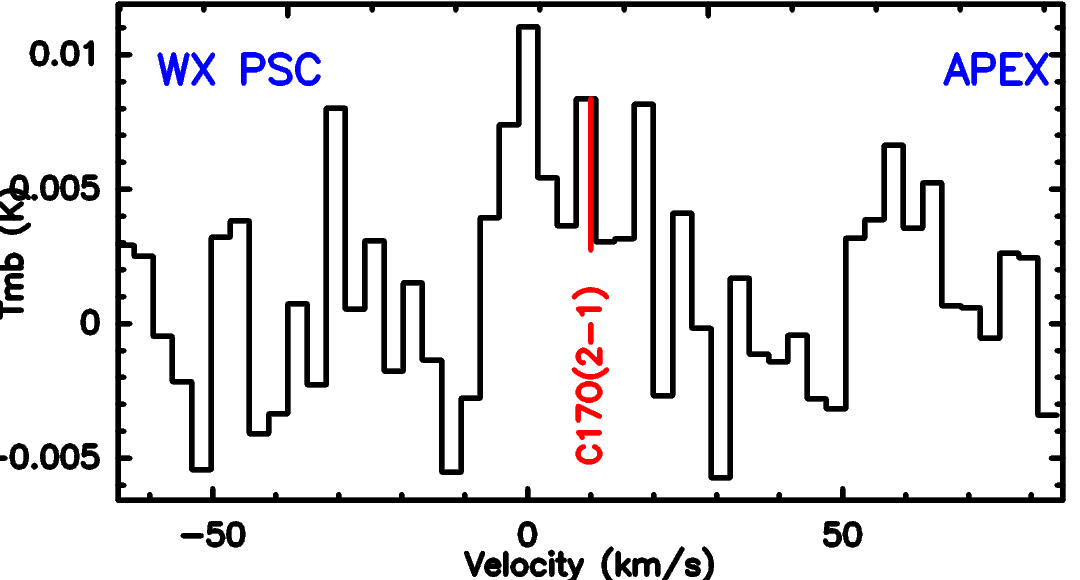} &
\includegraphics[width=.41\textwidth]{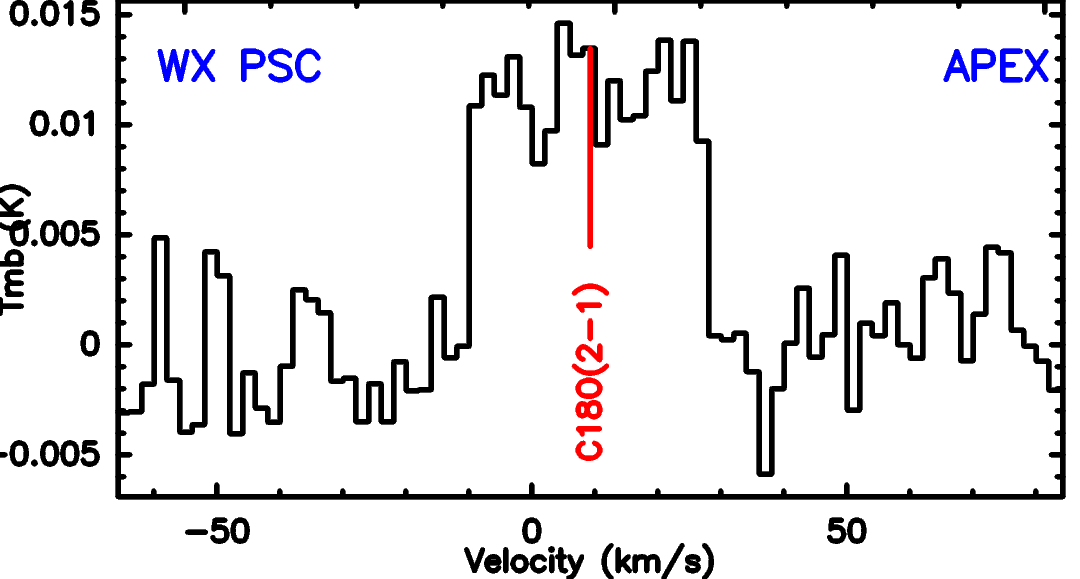} \\
\includegraphics[width=.41\textwidth]{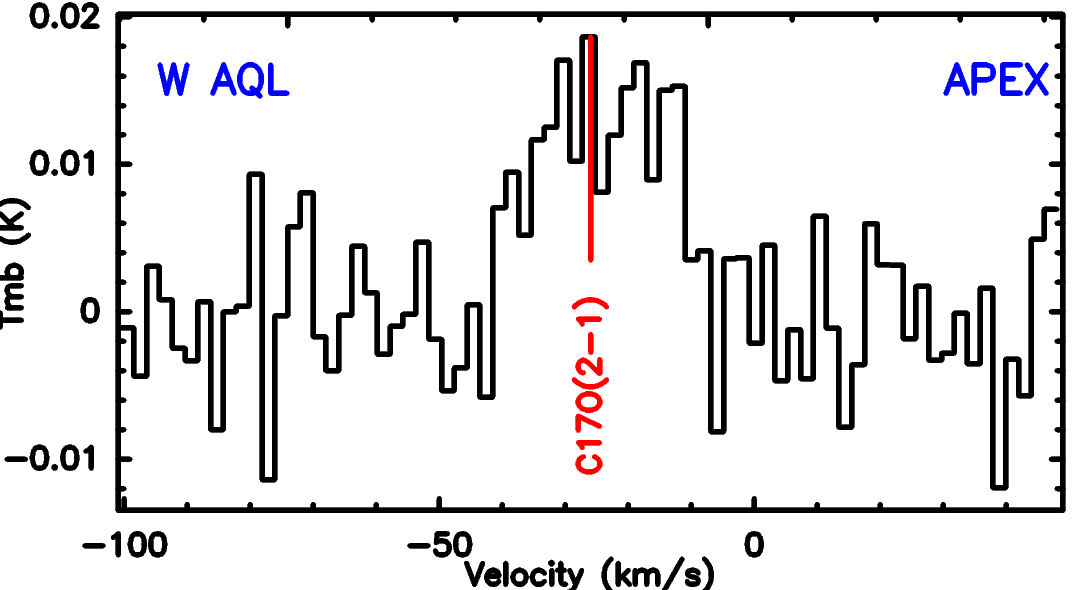} &
\includegraphics[width=.41\textwidth]{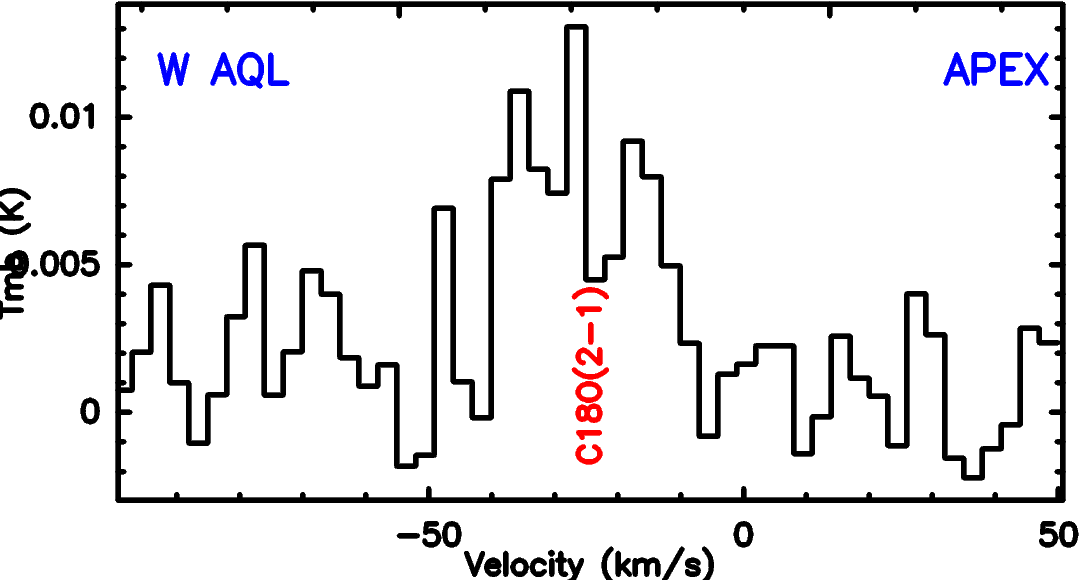} \\
\includegraphics[width=.41\textwidth]{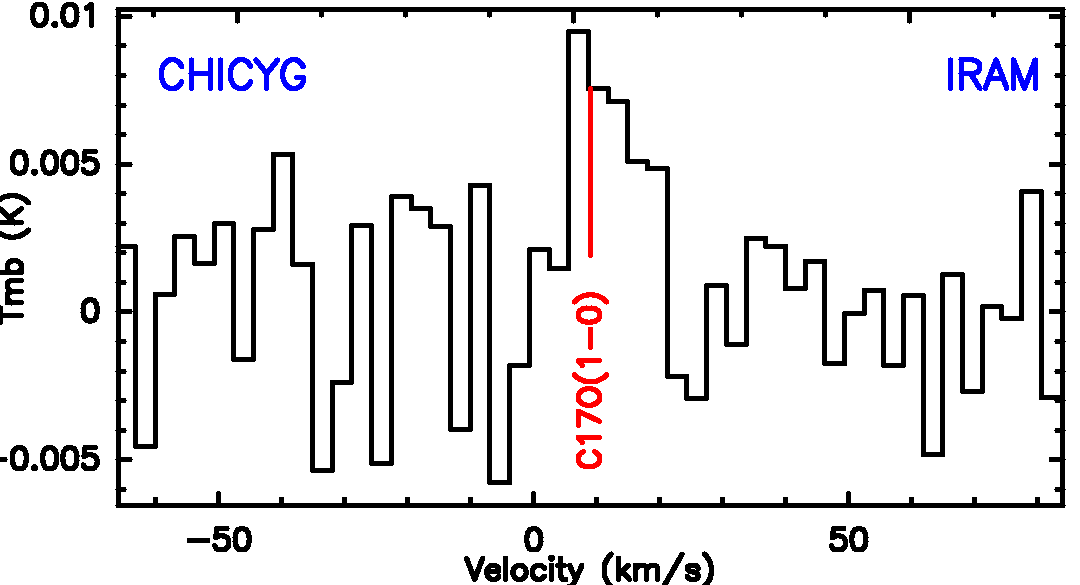} &
\includegraphics[width=.41\textwidth]{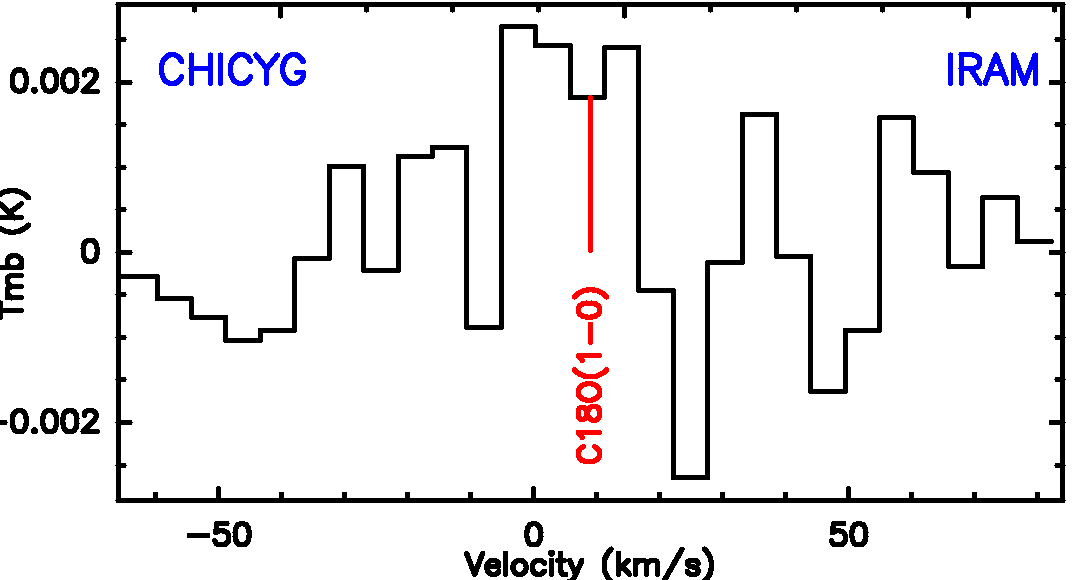} \\
\includegraphics[width=.41\textwidth]{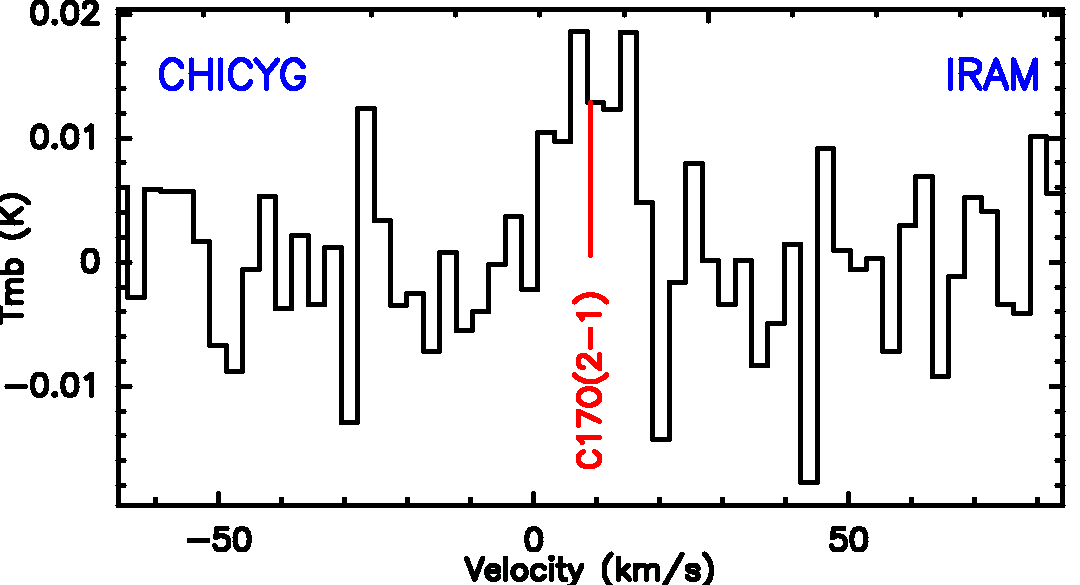} &
\includegraphics[width=.41\textwidth]{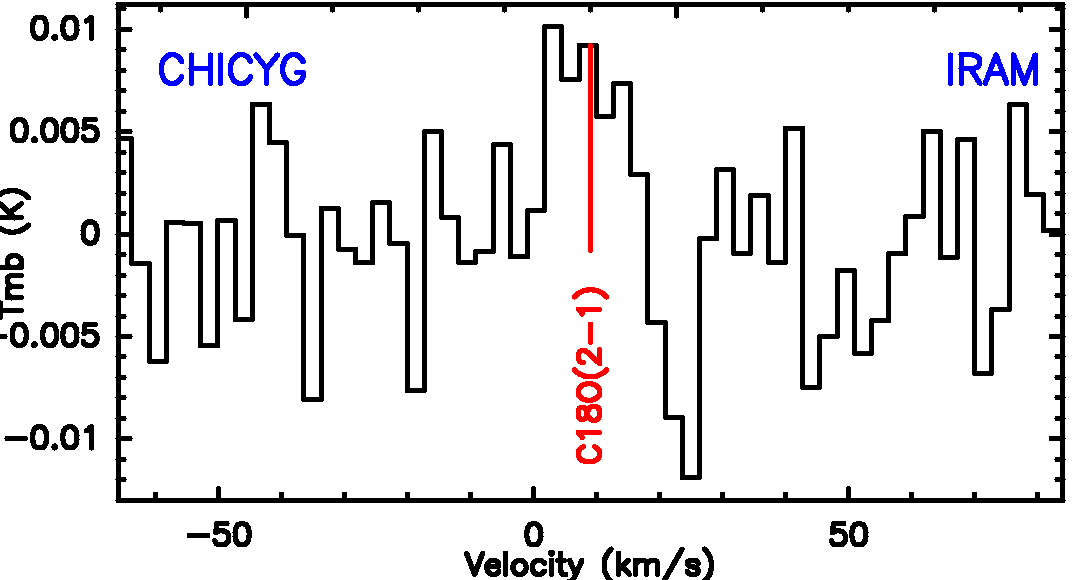} \\
\end{tabular}
\caption{CO isotopologue line detections, plotted in main-beam-temperature scale $T_{mb}$.}
\label{fig:profiles}
\end{figure*}

\begin{figure*}[!htb]
\ContinuedFloat
\centering
\begin{tabular}{cc}
\includegraphics[width=.41\textwidth]{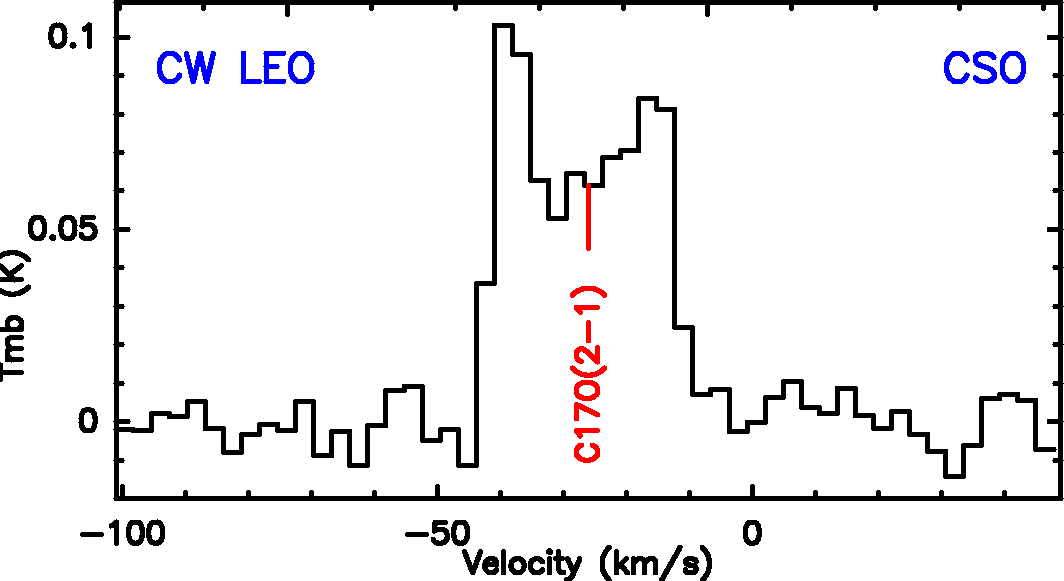} &
\includegraphics[width=.41\textwidth]{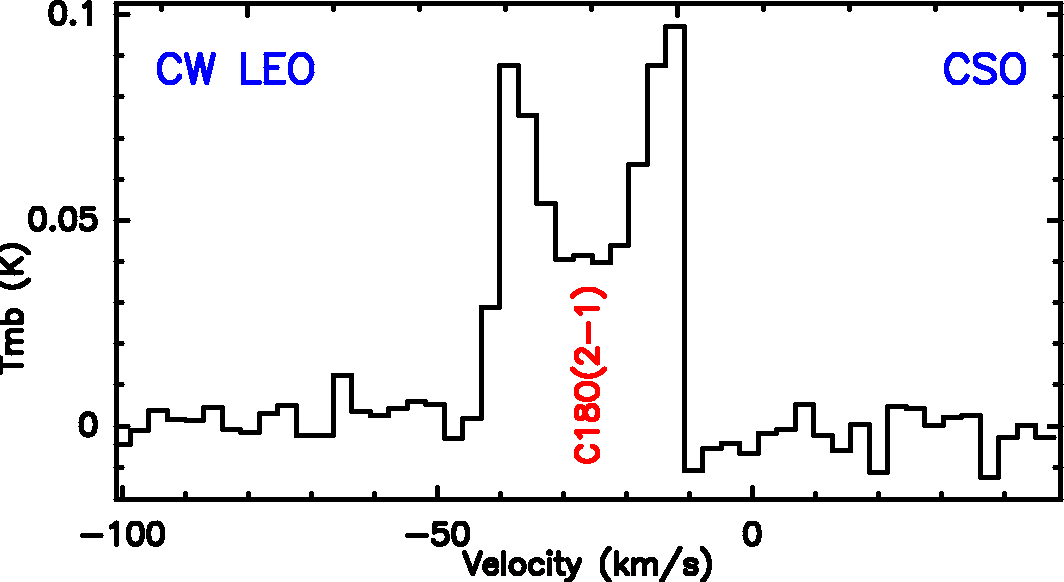} \\
\includegraphics[width=.41\textwidth]{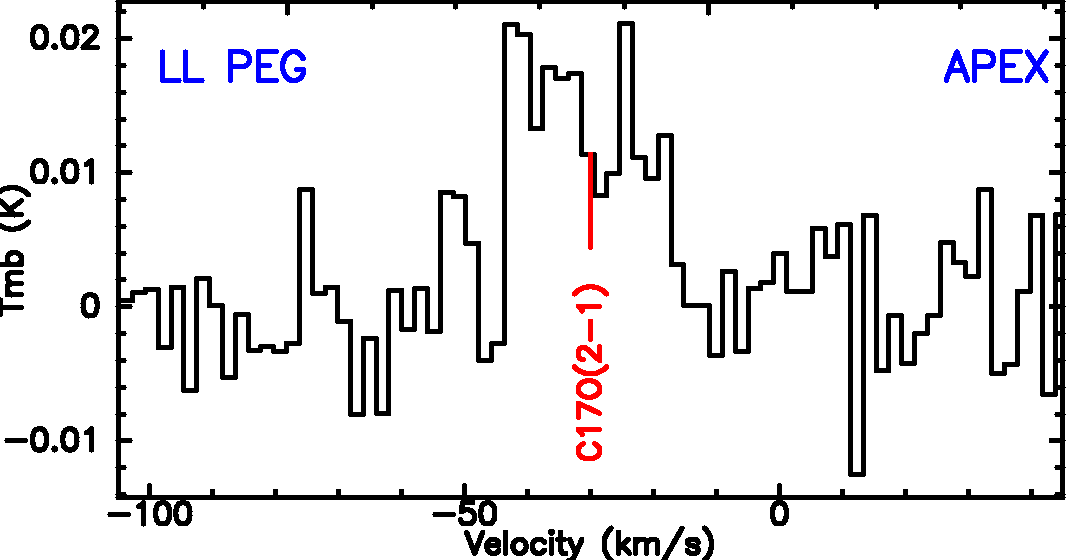} &
\includegraphics[width=.41\textwidth]{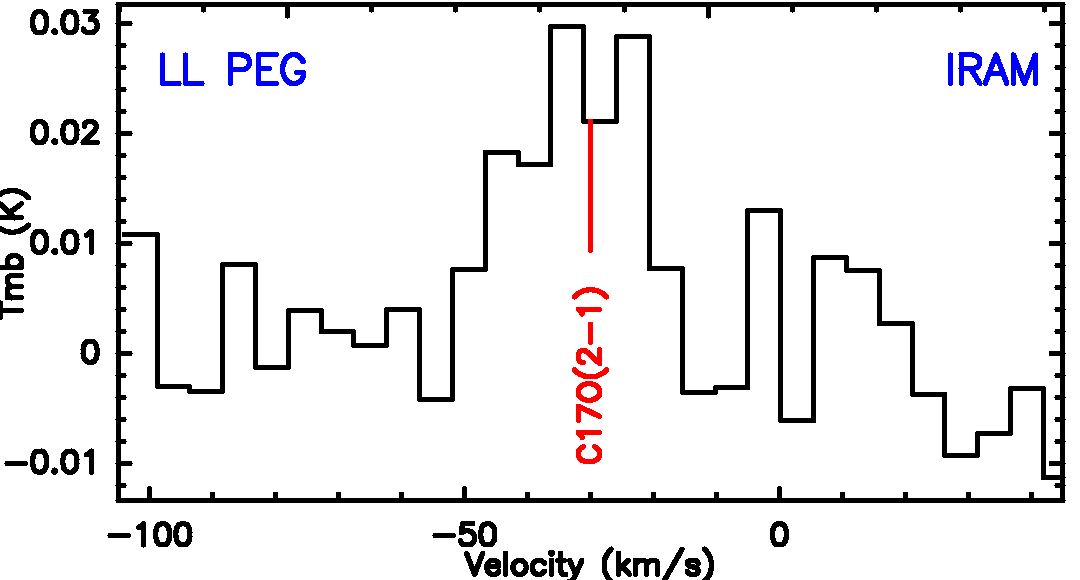} \\
\includegraphics[width=.41\textwidth]{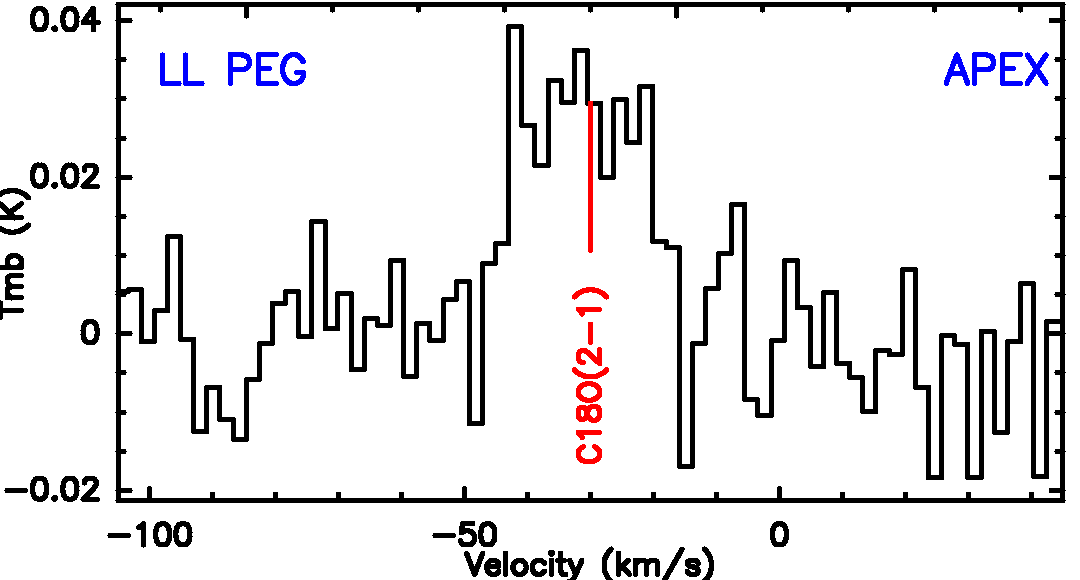} &
\includegraphics[width=.41\textwidth]{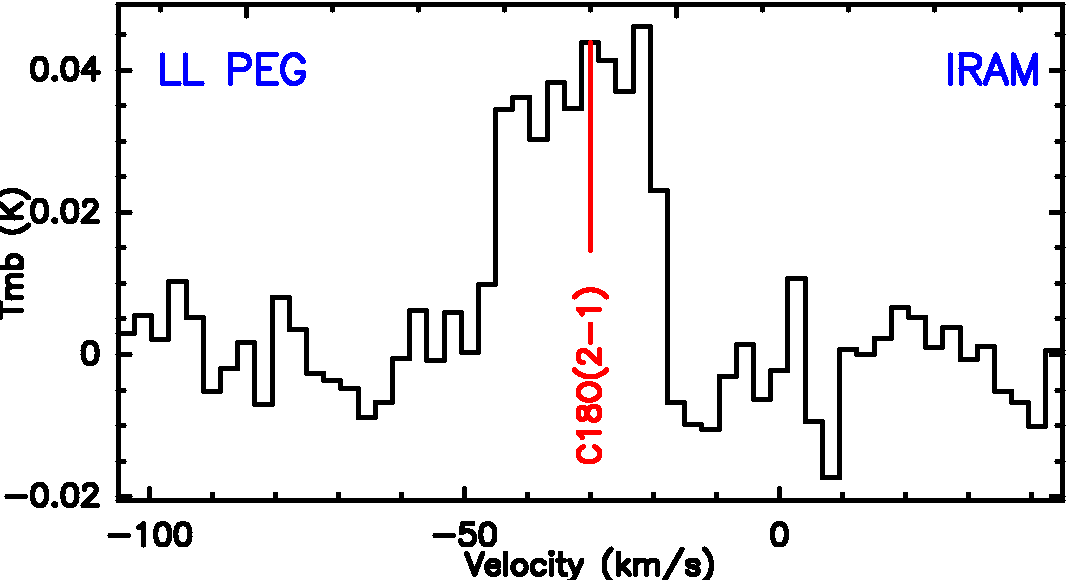} \\
\includegraphics[width=.41\textwidth]{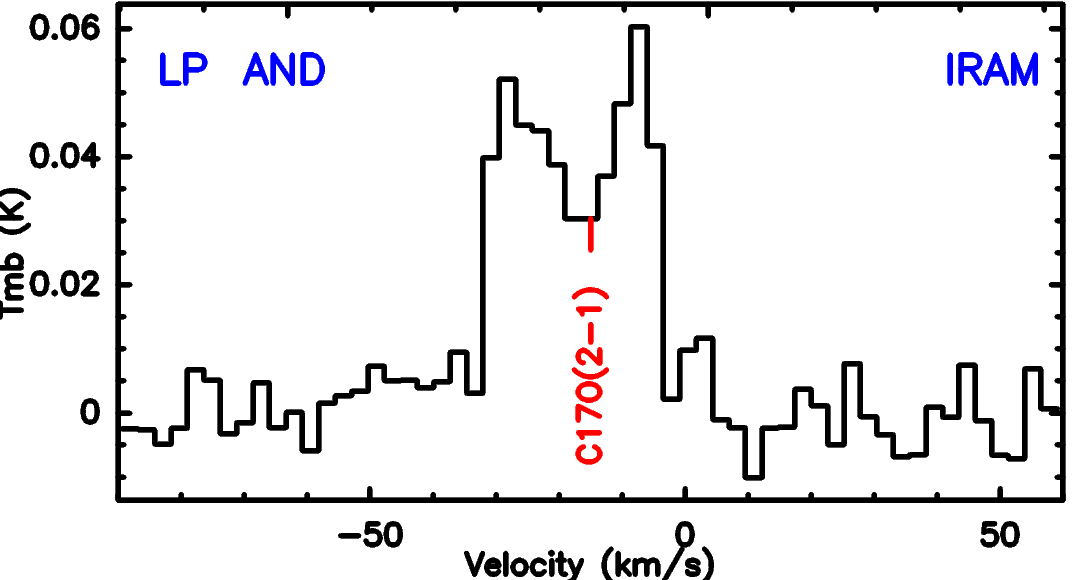} &
\includegraphics[width=.41\textwidth]{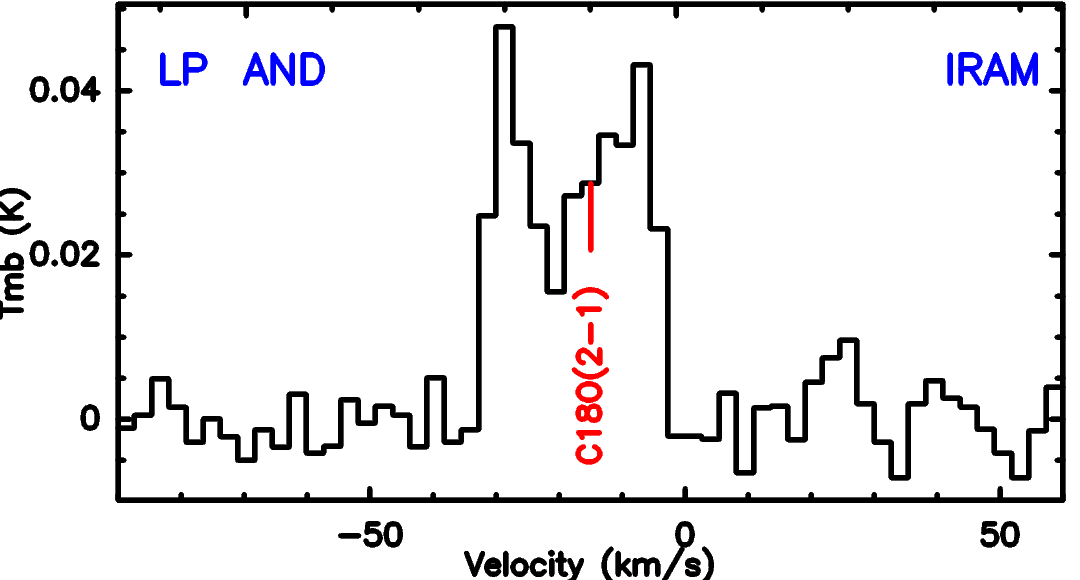} \\
\includegraphics[width=.41\textwidth]{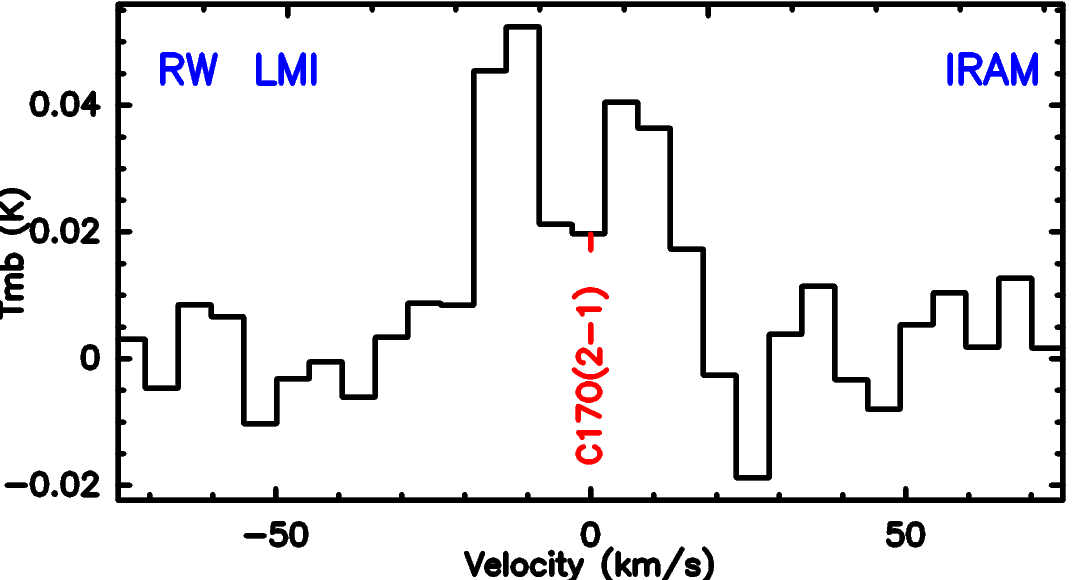} &
\includegraphics[width=.41\textwidth]{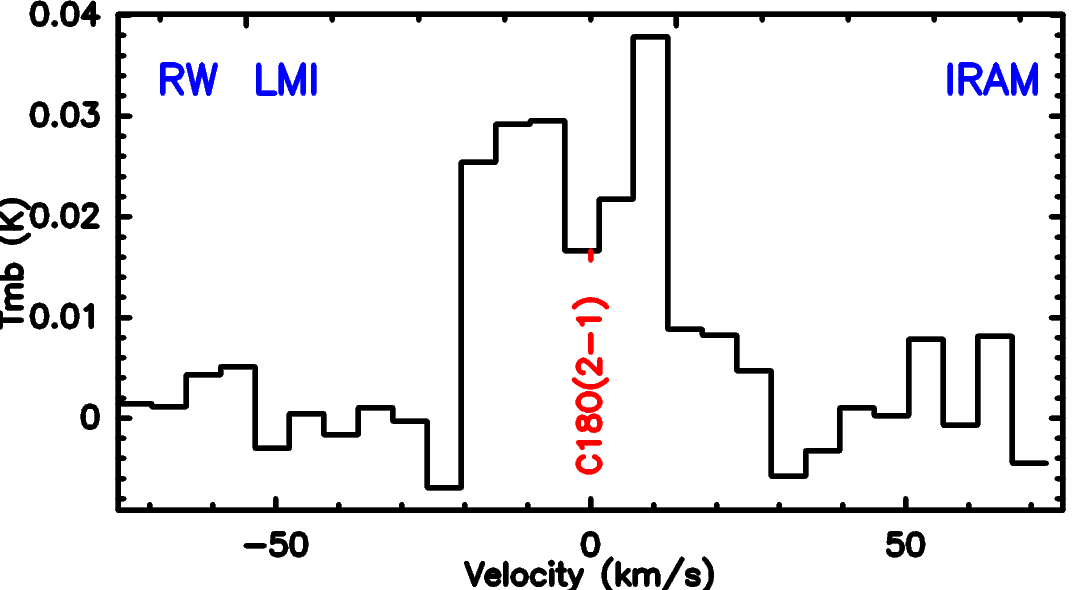} \\
\includegraphics[width=.41\textwidth]{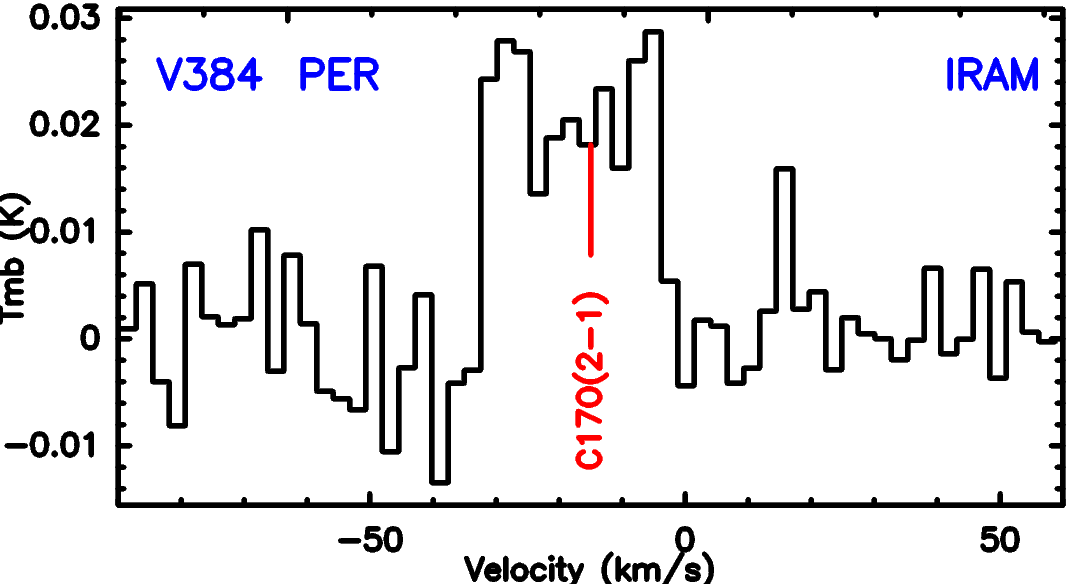} &
\includegraphics[width=.41\textwidth]{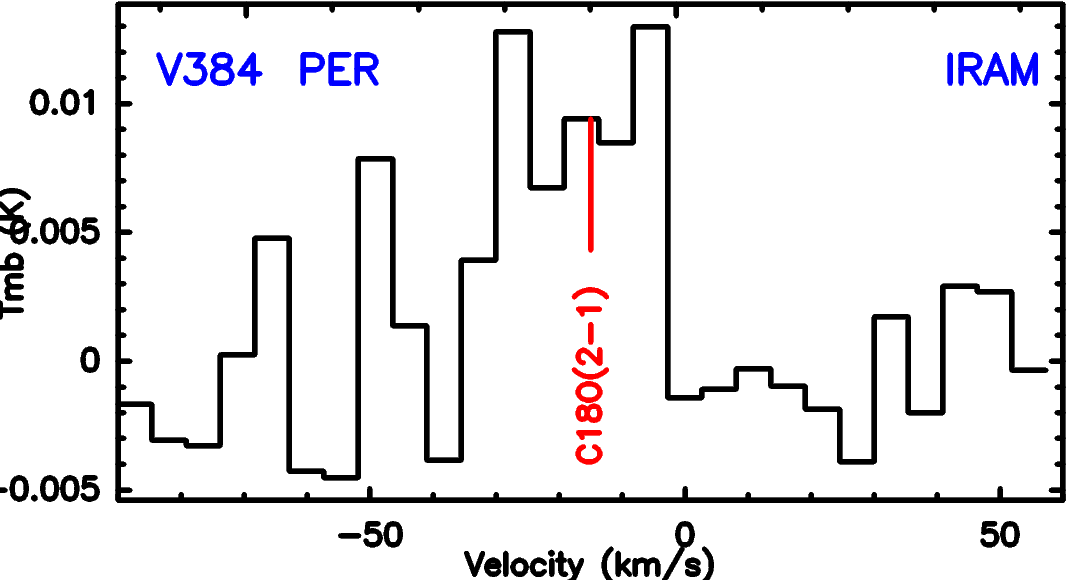} \\
\end{tabular}
\caption{Continued.}
\label{fig:profiles2}
\end{figure*}

\end{appendix}

\end{document}